# Vertical $\beta$-$Ga_2O_3$ Isolated Source Electrode Field Effect Transistors (ISEFET) Without Planarization or Mid-Gap Acceptor Blocking Layers

Akilesh Srikanth[1*], Md Saklain Morshed[1], Chandan Joishi[1], Ahmad E. Islam[3], and Siddharth Rajan[1,2]

[1]*Department of Electrical and Computer Engineering, The Ohio State University, Columbus, OH 43210, USA*

[2]*Department of Material Science and Engineering, The Ohio State University, Columbus, OH 43210, USA*

[3]*Air Force Research Laboratory, Sensors Directorate, Wright-Patterson AFB, Dayton, OH, 45433, USA*

*Corresponding author email: srikanth.24@osu.edu*

**Abstract:**

We propose and demonstrate the first vertical $\beta$-$Ga_2O_3$ device architecture without the use of planarization etch back processes or mid-gap acceptor regions. The Isolated Source Electrode Field Effect Transistor (ISEFET) incorporates a dielectric blocking layer to access an isolated source pad extending from the top fin metal. Scaled multi-fin channels were formed by electron beam lithography with a width of 200 nm along with the source pads and then etched to a trench depth of ~1.2 µm. The fabricated devices showed enhancement mode operation with threshold voltage of 2 V and on-off ratio > $10^7$ with excellent gate modulation characteristics. The resulting device proved to be comparable to existing vertical transistors and suitable for high-throughput prototyping and large-scale manufacturing of future Gallium oxide and other wide bandgap semiconductor devices.

**Introduction:**

Beta-Gallium oxide ($\beta$-$Ga_2O_3$) based devices have shown to be potential candidates for future power electronics applications due to their attractive material properties with ultrawide bandgap of $E_g$ ~ 4.8 eV and theoretical breakdown field up to 8 MV/cm [1-7]. This translates to a superior Baliga Figure of Merit (BFOM) than Si, SiC, and GaN and the availability of high-quality native melt-grown substrates makes it suitable for commercial production with high reliability and lower defects and cost. However, the absence of *p*-type doping in $\beta$-$Ga_2O_3$ [8-10] poses significant challenges to efficient device development. Vertical device architectures are preferred for power devices [11-15] compared to lateral counterparts [16-18] due to their high breakdown voltage and high current densities for large-area/current scaling. Successful demonstrations of vertical device topologies in $\beta$-$Ga_2O_3$ include the Trench MOSFET/FinFET [19-26], U-MOSFET [27-32], and CAVET [33-38]. Vertical FinFETs require multiple dielectric/metal planarization and precision etch processes to successfully produce a working device. U-MOSFET/CAVETs require current blocking layers (CBLs) using mid-gap acceptor regions commonly with N and Mg doping/ion implantation [39] [40] which degrade carrier transport and lead to dispersion and trapping effects. Recent advancements in vertical $\beta$-$Ga_2O_3$ FinFETs with scaled fin-shaped channels have shown superior device performance with high multi-kilovolt breakdown voltages and low on-resistance [41].

In this work, we report the successful demonstration of the vertical $\beta$-$Ga_2O_3$ Isolated Source Electrode Field Effect Transistor (ISEFET) made without complex fabrication steps by incorporating a dielectric current blocking layer (CBL) outside the active region of the device to access the source contact. The 3D device schematic for a single-fin ISEFET is shown in Fig 1. (a) and the lateral cross-section along the long dimension of the fin is shown in (b). The source fin metal is buried underneath the gate oxide and gate metal which are conformally deposited over the fins without any planarization etch back involved. This source metal is extended from one side onto a larger bond pad area on top of the CBL region where it can then be accessed. The entire length of the fin metal is defined to be continuous across both the CBL and the $\beta$-$Ga_2O_3$ surface. There is slight overlap of the gate with the part of the fin having the CBL underneath to ensure all parts of the channel can be electrostatically controlled and prevent any potential leakage paths at the edge of the CBL. This is also beneficial considering there is no need for any critical alignment steps to place the gate immediately at the very edge of the CBL. The source pad region is opened few microns away from the gate edge to prevent shorting of these terminals.

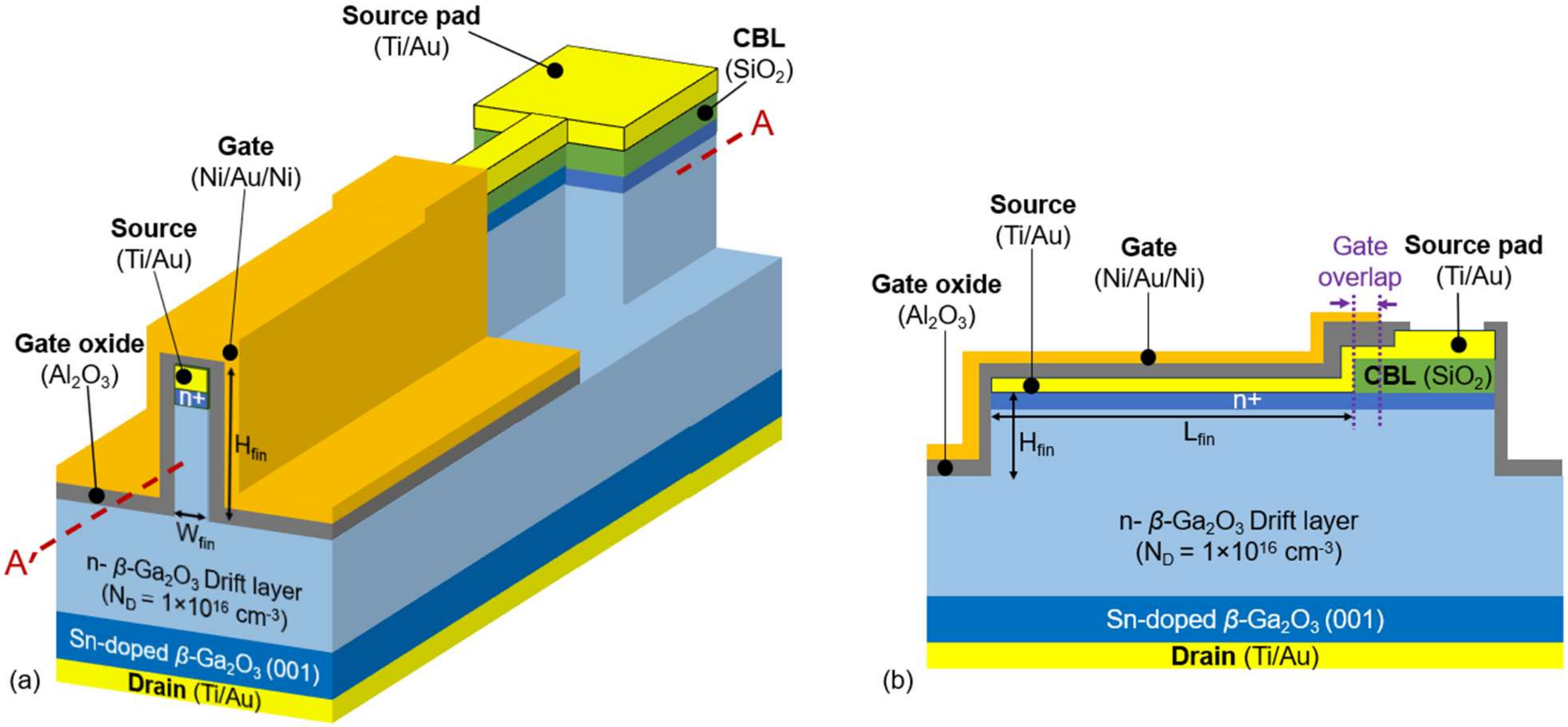

**Fig. 1.** (a) Three-dimensional schematic of a single-fin vertical ISEFET and (b) 2-D cross-section showing each layer along the A-A' plane

**Device Fabrication:**

The ISEFET devices reported here were fabricated on HVPE-grown (001) $\beta$-$Ga_2O_3$ Sn-doped conductive substrates obtained from Novel Crystal Technologies with a drift region thickness of 10 µm and $n^-$ doping of $1\times10^{16}$ cm$^{-3}$. The entire device process flow diagram is given in Fig. 2. The top $n$+ source contact layer with target depth of 150 nm and concentration of $5\times10^{19}$ cm$^{-3}$ [42] was formed through a multi-energy Si ion implantation. A uniform box plot profile was simulated using the Stopping Range of Ions in Matter (SRIM) software with energies ranging from 15 keV to 170 keV with a total dose of $9.16\times10^{14}$ atoms-cm$^{-2}$ as shown in Fig. 3. Post-implant activation annealing was done at 950ºC for 30 mins in $N_2$ at a reduced pressure of 360 Torr. Following this, 150 nm of $SiO_2$ was deposited using Plasma-Enhanced Chemical Vapor Deposition (PECVD) to serve as the current blocking layer (CBL) of the device. This was formed into separate regions using a combination of Inductively Coupled Plasma Reactive Ion Etching (ICP/RIE) and diluted Buffered HF wet etch. Various fin widths ranging from 0.2 μm-0.4 μm were patterned using Electron Beam Lithography in such a way that a small portion of the fins rested on top of the CBL. Fin metallization was done by electron-beam evaporation of Ti/Au/Ni (40 nm/100 nm/200 nm) to serve as both the source electrode and fin etch mask. An additional metallization step with the same metal stack was done on top of the CBL to form the source pad region. Self-aligned fin etching was done in a $BCl_3/Cl_2$-based ICP/RIE process to achieve a target depth of 1.2 μm using the top Ni as the etch mask. The sidewall etch damage was treated in <100ºC (85%) $H_3PO_4$ for 10 mins. Subsequently, the gate oxide was formed using Plasma-Enhanced Atomic Layer Deposition (PE-ALD) of 40 nm of $Al_2O_3$. Then, the gate metal was formed by e-beam evaporation and DC sputtering of Ni/Au/Ni (30 nm/70 nm/40 nm) for fin wrapping. Finally, the source pad was opened by removing the $Al_2O_3$ by RIE and drain metallization of Ti/Au (30 nm/100 nm) was done at the backside of the sample. After the source and drain Ohmic contact deposition, a post-metallization anneal (PMA) was performed at 470ºC for 1 min in $N_2$ ambient conditions [43].

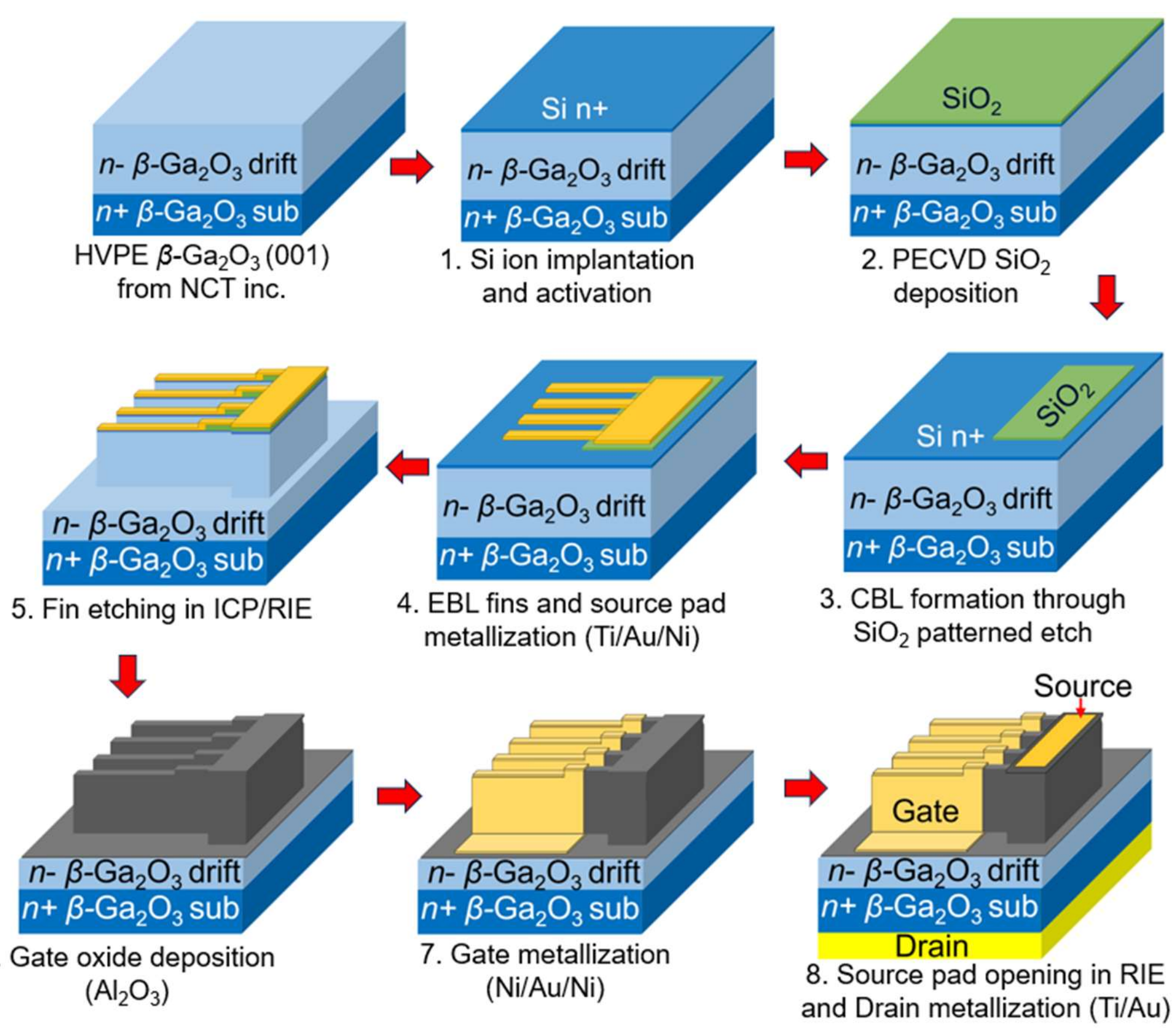

**Fig. 2.** Fabrication process flow for the multi-fin vertical ISEFET

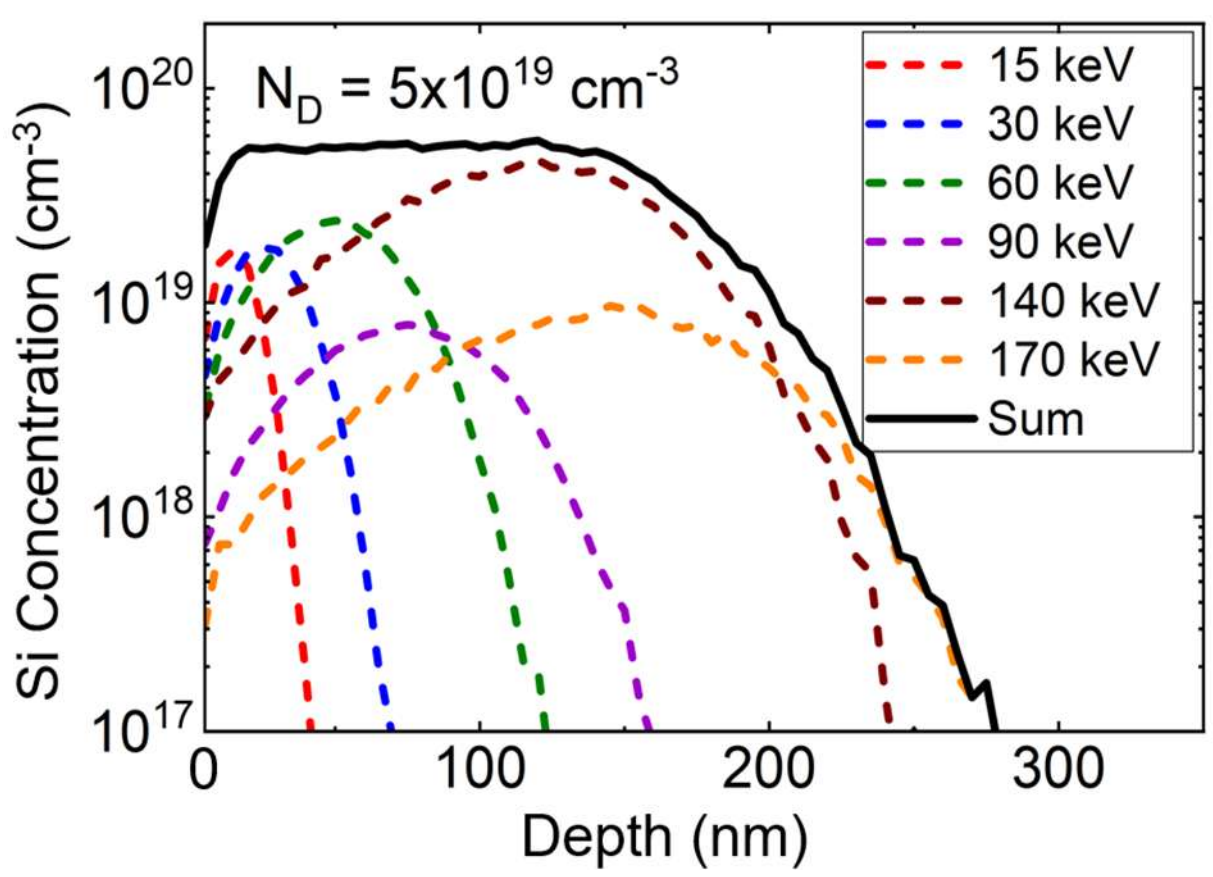

**Fig. 3.** Simulated SRIM Si implantation profile for n+ contact layer in $\beta$-$Ga_2O_3$

The completed 0.2 μm multi-fin ISEFET device top-down view is shown in Fig. 4 under optical microscope in (a) along with 80º tilted SEM images in (b) and (c). The fins were oriented along the [010] direction with (100)-like sidewalls as these are known to suffer less from the plasma etch damage [44] [45]. All the regions around the device have rounded corners to minimize electric field crowding and prevent pre-mature breakdown.

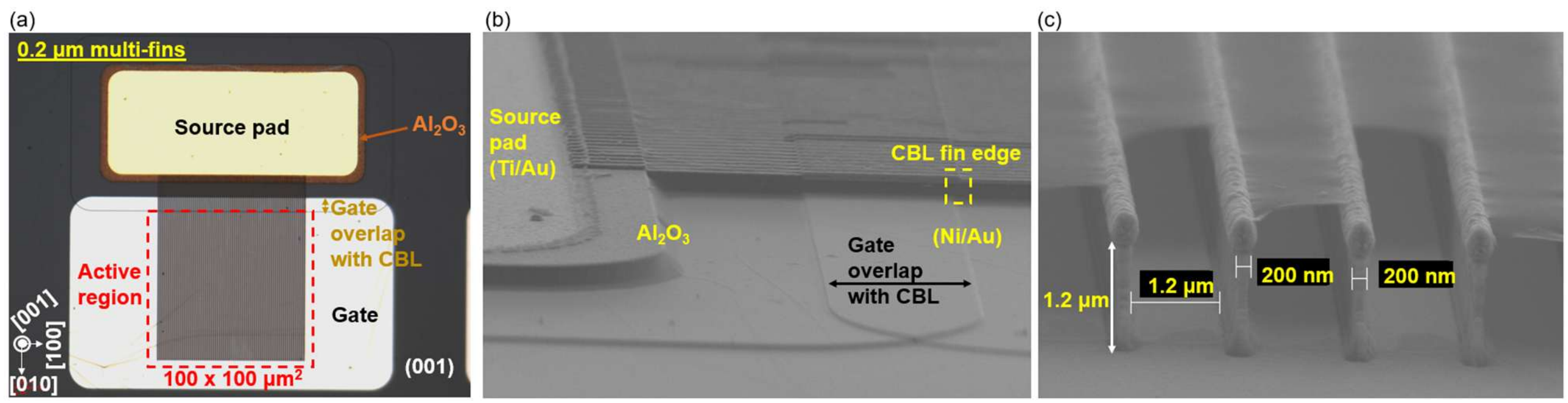

**Fig. 4.** Fully fabricated ISEFET 0.2 μm multi-fin device (a) top-view optical image and 80° tilted SEM image of fins: (b) lateral view and (c) vertical view

**Results and Discussions:**

All the testing done on this sample was measured using the Keysight B1505A Semiconductor Device Parameter Analyzer system. The results for this multi-fin device are normalized to an area of 100×100 μm$^2$ with measured fin width $W_{fin}$ = 0.2 μm, interfin spacing $S_{fin}$ = 1.2 μm, fin height $H_{fin}$ = 1.2 μm, and fin length $L_{fin}$ = 100 μm. These device dimensions have the highest aspect ratio corresponding to $L_g$:$W_{fin}$ = 12:1 with double-sided gated sidewalls. The transfer characteristics ($J_D$-$V_{GS}$) for the ISEFET at a positive drain bias ($V_{DS}$ = 1 V) is shown in Fig. 5. (a). The threshold voltage was extracted to be $V_{Th}$ = 2 V which shows the fins are fully depleted at zero bias and therefore can be operated in E-mode. This device also demonstrates a high $I_{on}/I_{off}$ ratio > $10^7$ with very low gate leakage ($J_G$) observed. There is significant hysteresis observed in these devices after sweeping the gate ($V_{GS}$) from depletion bias of -3 V to accumulation at 10 V and back potentially due to the large trap $D_{it}$ at the $Al_2O_3/\beta$-$Ga_2O_3$ interface from plasma etch damage. This points out that the specific $H_3PO_4$ treatment method used for these devices was insufficient to promote a clean growth surface prior to ALD $Al_2O_3$ deposition. Gate dielectric stacks showing low hysteresis have been demonstrated for Gallium Oxide structures [46-48], and this technology can be adopted for future improvements to the proposed ISEFET.

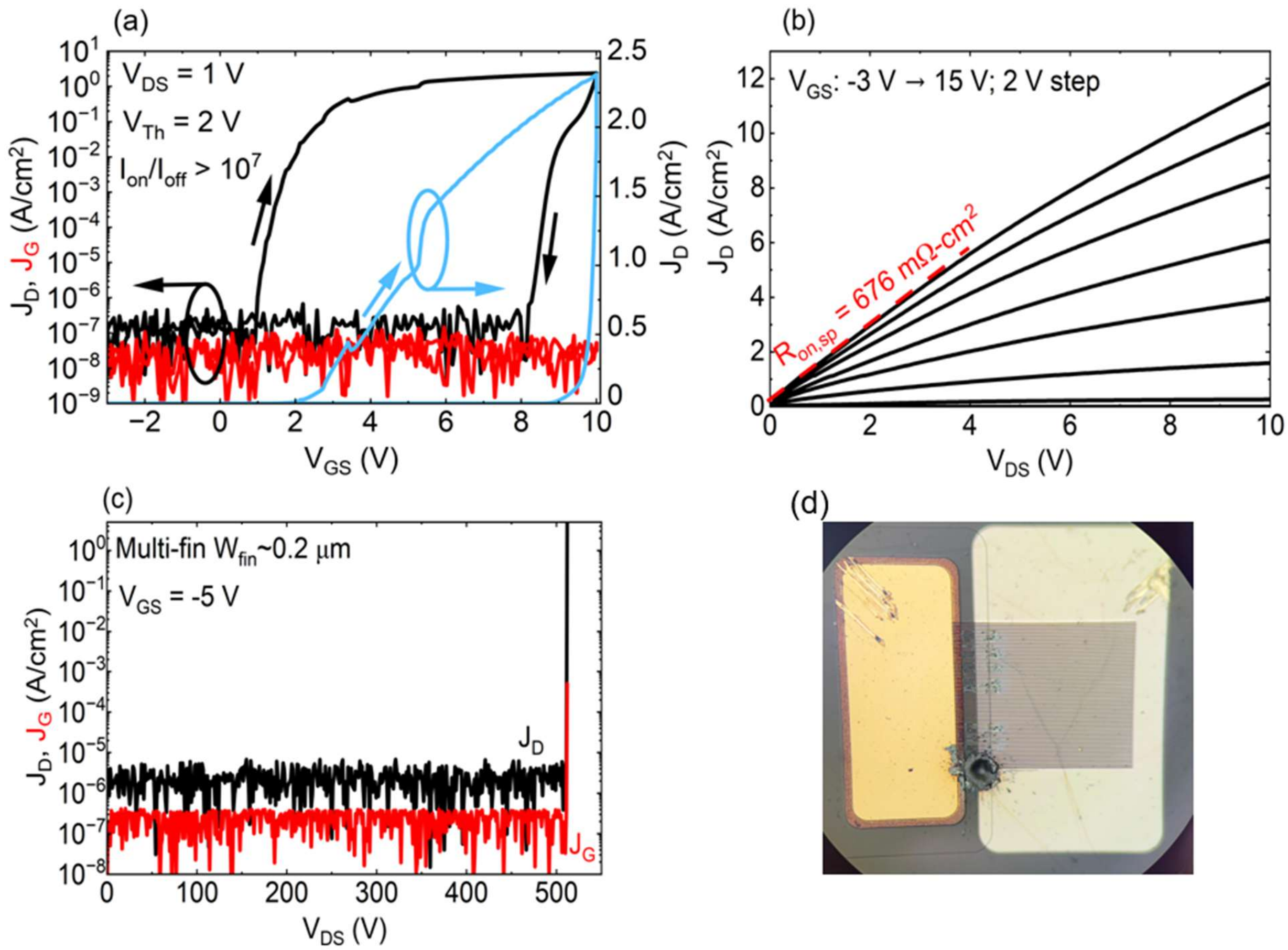

**Fig. 5.** (a) DC transfer, (b) output, and (c) off-state breakdown characteristics of the fabricated $\beta$-$Ga_2O_3$ ISEFET. (d) Optical micrograph of the device showing destructive breakdown at fin edge

The DC output characteristics ($J_D$-$V_{DS}$) for the ISEFET with the same dimensions as earlier is shown in Fig. 5. (b). The device has excellent gate modulation with an extracted on-resistance $R_{on\text{-}sp}$ = 676 m$\Omega$-cm$^2$ at an applied gate bias $V_{GS}$ = 15 V. This relatively high resistance was later attributed to the impartial activation/degradation of the Si-implanted n+ contact layer for this particular sample. Due to this, the highest attained output current density for this device was 12 A/cm$^2$ at drain bias $V_{DS}$ = 10 V and $V_{GS}$ = 15 V. Additionally, relatively lower aspect ratio structures $L_g$:$W_{fin}$ = 6:1 with $W_{fin}$ = 0.4 μm was also characterized (not shown here). These devices had a maximum output current density of 70 A/cm$^2$ at $V_{DS}$ = 10 V and $V_{GS}$ = 15 V, $I_{on}/I_{off}$ ratio > $10^9$ at $V_{DS}$ = 10 V, and $R_{on-sp}$ = 322 mΩ-cm$^2$ which is 2x lower than the previous device with $W_{fin}$ = 0.2 μm. However, it showed triode-like output characteristics with no saturation regime at high drain bias.

Fig. 5. (c) shows the off-state breakdown characteristics for the multi-fin ($W_{fin}$ = 0.2 μm) ISEFET at $V_{GS}$ = -5 V. The observed breakdown voltage was around $V_{DS}$ = 510 V while maintaining a low leakage current ($J_G \sim 10^{-7}$ A/cm$^2$) at the noise floor of the system throughout the measurement until the breakdown point. This corresponds to a non-punch through breakdown with an electric field of $E_c$ = 1.34 MV/cm in the $\beta$-$Ga_2O_3$ drift region. The physical location of the breakdown was seen at the gate edge of the device periphery close to the source pad region as shown in the optical micrograph in Fig. 5. (d). Prior measurements of the dielectric strength of this specific process of ALD $Al_2O_3$ have indicated similar breakdown values so the $V_{BR}$ of the transistor is most likely limited by the gate oxide at the MOS interface. Effective field management strategies and oxide spacers at the trench bottom are potential future design criteria for the next generation of $\beta$-$Ga_2O_3$ ISEFET devices to further improve the $V_{BR}$.

**Conclusion:**

In summary, this work shows the fabrication process and results for the first generation of vertical $\beta$-$Ga_2O_3$ ISEFET achieved without planarization etch back steps or mid-gap acceptor CBL regions by incorporating an isolated source electrode to serve as a large area source contact to the fin metal. This device design allows for rapid prototyping and high yield innovative ideas that could be incorporated into future iterations of the ISEFET without being limited by the nature of the manufacturing process. Initial results for a 0.2 μm multi-fin device showed $V_{Th}$ = 2 V, $I_{on}/I_{off}$ > $10^7$, $R_{on-sp}$ = 676 mΩ-cm$^2$, and $V_{BR}$ = 510 V. This device exhibits transistor characteristics that are not inherently limited by the device architecture itself and rather requires optimization of specific process steps for further improvements. This provides a promising path for development of highly manufacturable vertical $\beta$-$Ga_2O_3$ high-voltage transistors for power electronics applications.

**Acknowledgements:**

A major portion of this work was performed in Nanotech West Cleanroom facility and Characterization Lab. We acknowledge the funding support from the Defense Associated Graduate Student Innovators (DAGSI) Fellowship.